\documentclass[final]{svjour3}
\usepackage{graphicx}
\usepackage{rotating}
\usepackage{amssymb}
\usepackage{mathptmx}

\usepackage[numbers]{natbib}
\makeatletter
\journalname{Journal of Low Temperature Physics}

\begin{document}

\newcommand{\umux}{$\mu$MUX }
\newcommand{\sups}[1]{\textsuperscript{#1}}
\newcommand{\hdblarrow}{H\makebox[0.9ex][l]{$\downdownarrows$}-}

\title{Characterization of aliased noise in the  Advanced ACTPol receiver}

\author{P. A. Gallardo\sups{1}\and 
        M. D. Niemack\sups{1}\and 
        J. E. Austermann \sups{2}\and
        J. Beall\sups{2}\and
        N. F. Cothard\sups{1}\and
        C. J. Duell\sups{1}\and
        S. M. Duff\sups{2}\and
        S. W. Henderson\sups{3}\sups{4}\and
        G. C. Hilton\sups{2}\and
        S. P. Ho\sups{5}\and
        J. Hubmayr\sups{2}\and
        C. D. Reintsema\sups{2}\and
        M. Salatino\sups{6}\sups{7}\and
        J. N. Ullom\sups{2}\and
        J. Van Lanen \sups{2}
        M. R. Vissers\sups{2}\and
        E. J. Wollack\sups{8}
        }

\institute{
\sups{1} Department of Physics, Cornell University, Ithaca, NY 14850, USA\\ \email{pag227@cornell.edu} \\
\sups{2} National Institute of Standards and Technology, 325 Broadway, Boulder, CO 80305, USA \\
\sups{3} Kavli Institute for Particle Astrophysics and Cosmology, Menlo Park, CA 94025, USA\\
\sups{4} SLAC National Accelerator Laboratory, Menlo Park, CA 94025, USA\\
\sups{5} Department of Physics, Princeton University, Princeton, NJ 08540, USA \\
\sups{6} Stanford University, 382 via Pueblo, 94305 Stanford CA, USA \\
\sups{7} Kavli Institute for Particle Astrophysics and Cosmology, 452 Lomita Mall,  94305  Stanford CA, USA \\
\sups{8} Goddard Space Flight Center, National Aeronautics and Space Administration, Greenbelt,MD 20771, USA
}

\maketitle

\begin{abstract}

Advanced ACTPol is the second generation polarization-sensitive upgrade to the $6\, \rm m$ aperture Atacama Cosmology Telescope (ACT), which increased detector count and frequency coverage compared to the previous ACTPol receiver. Advanced ACTPol utilizes a new two-stage time-division multiplexing readout architecture based on superconducting quantum interference devices (SQUIDs) to achieve a multiplexing factor as high as 64 (rows), fielding a 2,012 detector camera at 150/220 GHz and two 90/150 GHz cameras containing 1,716 detectors each. In a time domain system, aliasing introduces noise to the readout. In this work we present a figure of merit to measure this noise contribution and present measurements of the  aliased noise fraction of the Advanced ACTPol receiver as deployed.

\keywords{Aliasing, Aliased Noise, Time Division Multiplexing, Transition-Edge Sensor.}

\end{abstract}

\section{Introduction}
The Atacama Cosmology Telescope ACT is a six-meter telescope located in the Atacama desert intended to measure with high (arcminute) angular resolution the Cosmic Microwave Background (CMB) radiation. Advanced ACTPol, a third generation receiver for ACT is the second upgrade to its cryogenic camera, which fields three two kilo-pixel arrays of time-domain-multiplexed (linear) polarization-sensitive dichroic horn-coupled superconducting Transition Edge Sensors  (TESes)  \cite{henderson_advanced_2016, henderson_highly-multiplexed_2018}. The three arrays in Advanced ACTPol are named PA4, PA5 and PA6 and feature 64, 55 and 55 rows (and 32 columns) of detectors at 150-220, 90-150 and 90-150 GHz respectively.

The Advanced ACTPol array is read out using time domain multiplexing. Each column is read in the time domain by visiting its rows successively. Rows are sampled by the Multi Channel Electronics \cite{battistelli_functional_2008} (MCE), which  digitally (anti-alias) filters and decimates the read out feedback current on each TES to limit the bandwidth of the stored data stream. The user has access to this anti-aliased down-sampled version of the feedback current.

\begin{figure}
    \centering
    \includegraphics[width=\textwidth]{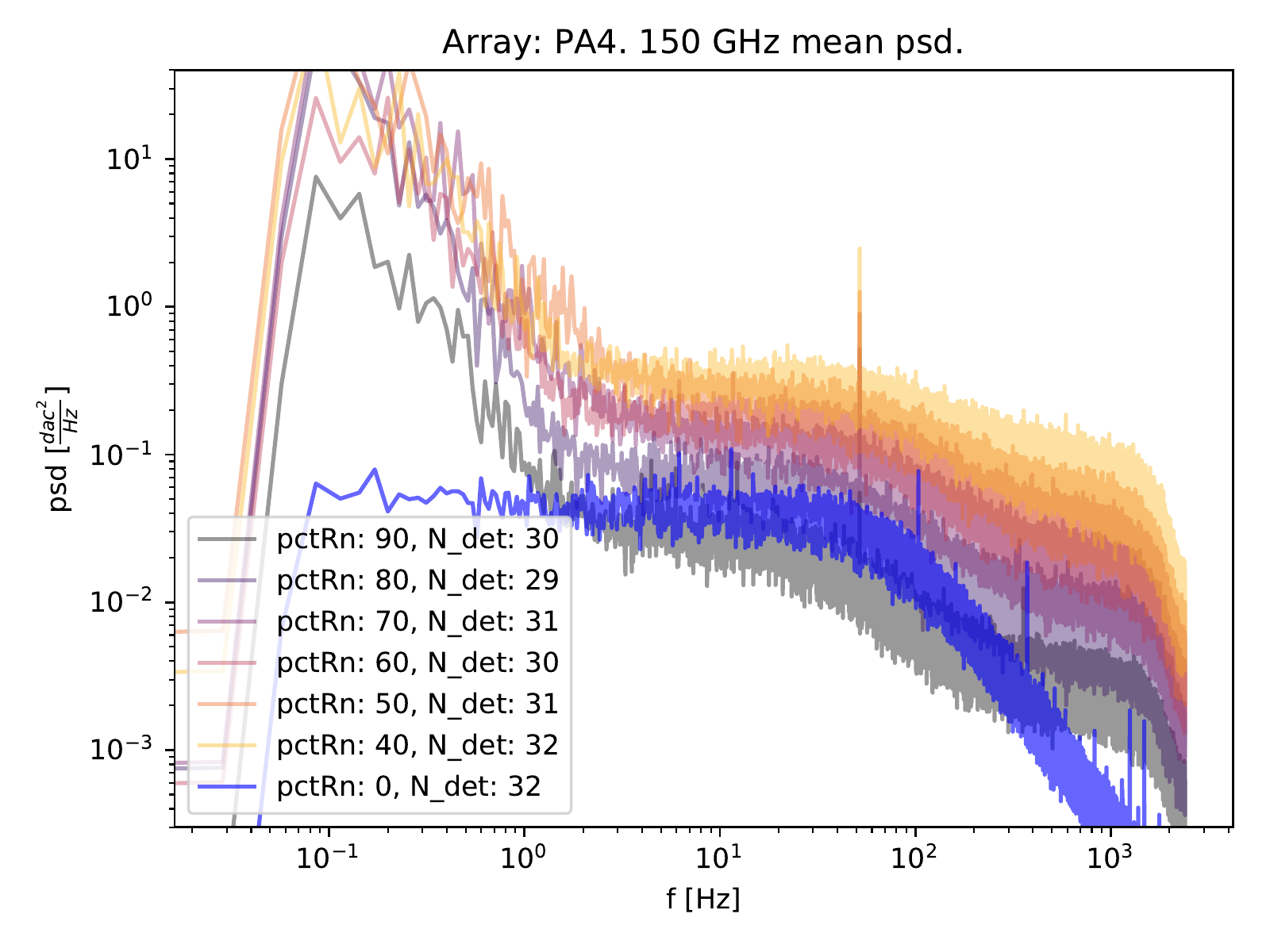}
    \caption{Average noise spectrum for $150\, \rm{GHz}$ detectors in PA4 acquired by reading out only 4 rows which gives a sampling frequency $16\times$ faster than nominal science observation mode (see table \ref{tab:Aliasing_mceAquisitionParams}). Detectors shown have been flagged as nominally operating using the aliasing fraction cut described in Section \ref{sec:Measurements}.}
    \label{fig:average_spectrum}
\end{figure}

Aliasing arises when a digitally sampled continuous-time signal has
frequency content above the Nyquist frequency ($f_{Ny}$, half of the sampling frequency $f_s$). 
Higher frequencies than $f_{Ny}$  fold to lower frequencies and
their power adds to the low frequency content \cite{oppenheim_signals_1997, gallardo_optimizing_2019}. This effect, inherent to the process of sampling adds the out-of-band power to the in-band power spectrum. In the case of a detector system, this out-of-band power adds to the inherent detector noise  increasing the apparent noise power spectral density by a fraction (see section \ref{sec:fom}), which depends on the details of the out-of-band noise. It is common (in the design stage) to add a hardware RL filter to mitigate the out-of-band power in a TES bolometer system, however there is a trade-off between removing frequency content (bandwidth) from the detector by filtering and the ability to bias it (stability). The way the hardware anti-aliasing filter operates with a particular type of sensor often is studied experimentally as the interaction is complicated in practice (it depends on the excess detector noise, readout noise, RL filter fall-off, etc). Figure \ref{fig:average_spectrum} shows the average noise power spectral density for varying bias points in the transition from 40 to $90\% \, \rm {R_n}$ (where $\rm{R_n}$ is the normal resistance of the device) on PA4 read out using only 4 rows, which gives a sampling frequency $16 $ times faster than the nominal science observation mode (see table \ref{tab:Aliasing_mceAquisitionParams} for details) allowing us to probe the high frequency part of the noise in the detection system.

In photon background-limited CMB experiments like Advanced ACTPol, the noise budget of the system is dominated by photon noise; the mapping speed of the overall experiment depends on the total noise budget. Percent level changes in this budget  decrease mapping speed \cite{hill_bolocalc:_2018}. Because aliasing occurs at the end of the detection chain, it is of interest to characterize how it impacts the overall noise budget.

In this work we present measurements of the aliased noise characteristics of the Advanced ACTPol receiver as it was deployed in the field. Measurements are done by re-configuring the software readout electronics to increase the sampling frequency at which detectors are read out at the expense of reading only a fraction of the array. In section \ref{sec:fom} we describe the figure of merit used in this study to measure the aliasing noise contribution. In section \ref{sec:Measurements} we describe the measurement strategy. In section \ref{sec:Results} we discuss our results and in section \ref{sec:Conclusion} we conclude and interpret our results.

\section{Figure of Merit}
\label{sec:fom}

We quantify aliased noise by evaluating the fraction defined as \begin{equation}
\label{eq:af}
    AliasFraction = \frac{\bar P_{slow} [dac^2/Hz]}{\bar P_{fast}[dac^2/Hz]},
\end{equation}
where $\bar P$ denotes an estimator for the mean power spectral density (in units of $dac^2/Hz$) within the 10 to 60 Hz band at the nominal observing sampling frequency (denoted slow) and a higher sampling frequency (denoted fast). In this study we use the median as an estimator for the mean for being a robust estimator in the presence of line contamination. The sampling frequencies (before decimation) that are accessed are $7.8\, \rm{kHz}$ ($9.1\,\rm{kHz}$) and $125 \, \rm {kHz}$ ($125\, \rm{kHz}$) for PA4 (PA5 and PA6).

This figure of merit has the benefit of being agnostic to modeling systematic errors (as it is purely observational) that can arise from modeling the detector and readout noise in the system.

\section{Measurements}
\label{sec:Measurements}


We reconfigure the Advanced ACTPol readout system \cite{henderson_readout_2016} (MCE) to sample at a faster than nominal science (CMB) observation sampling speeds. We achieve this by varying the number of rows the MCE visits in each column. By having less rows to read, the multiplexer spends less time reading out one column and therefore the overall sampling rate increases. The sampling rate in the MCE is given by
\begin{equation}
    \label{eq:fs}
    f_{s} = \frac{50\, \rm{MHz}}{num\_rows \times row\_len},
\end{equation}
where $num\_rows$ is the number of rows to be read and $row\_len$ is the number of $50\,\rm MHz$ cycles spent per row.
This digitally sampled signal is then digitally filtered via a Butterworth  filter  and a time-stream with sampling frequency 
\begin{equation}
    \label{eq:fs-readout}
    f_{s\_readout} = \frac{50\,\rm{MHz}}{num\_rows \times row\_len \times data\_rate}
\end{equation}
is given to the user, where $data\_rate$ is the decimation factor at which data are subsampled. In nominal science observation mode, the (decimated) sampling frequency returned to the user is 300.5 Hz for PA4 (64 rows) and 395 Hz for PA5 and PA6 (55 rows). In this study we read out down to 4 rows which allows to obtain decimated sampling frequencies as high as 4.8kHz for PA4 and 5.4kHz for PA5 and PA6. 

Data is acquired with reflective covers on the receiver window entrance. The MCE is configured to acquire data  at different sampling frequencies with the parameter $num\_{rows}$ varying between 4 and the total number of rows in the array. Table \ref{tab:Aliasing_mceAquisitionParams} shows the parameters used in the MCE and their corresponding sampling frequencies as well as the Butterworth filter frequencies at 0.99 gain. We acquire one minute long time streams in each acquisition.
\begin{table}
    \centering
    \begin{tabular}{c|c|c|c|c|c|c}
        array & num\_rows & row\_len & data\_rate & $f_s$[kHz] & $f_{s\_readout}$[Hz] & $f_{c}[Hz]$\\
        \hline \hline
         PA4      & 64, 4 & 100 & 26 & 7.8, 125 &  300.5, 4807 &  64.4, 1031.5\\
         PA5, PA6 & 55, 4 & 100 & 23 & 9.1, 125 &  395.3, 5435 & 75.0, 1031.5 
    \end{tabular}
    \caption{MCE parameters used to vary the sampling frequency. Sampling frequencies obtained (defined according to eqs. \ref{eq:fs} and \ref{eq:fs-readout}) are also shown. Digital anti-alias Butterworth filter cutoff frequency at $99\%$ gain is shown as $f_c$.}
    \label{tab:Aliasing_mceAquisitionParams}
\end{table}
Data was visually inspected to discard time intervals where glitches are present. We find two kinds of glitches in the data: discontinuities in the data and short pulse glitches. Discontinuities are found to occur when the data acquisition is started soon after one IV curve is acquired. We remedy this by waiting five minutes after each IV curve to allow the array to reach thermal equilibrium. The dataset presented here showed no identifiable discontinuities. We also find fast glitches, which are jumps in the readout that last one or a few samples in length. We interpret these jumps as cosmic ray hits and remove them from the data by selecting the longest glitch free data in each time stream on a detector to detector basis. One or two glitches (per array) in a one minute acquisition interval are typical in the data stream.

For each glitch-free section of data we compute the power spectral density (periodogram) via the Welch method \cite{scipy_2001}. We compute the median power spectral density in a band that goes from 10 to $60 \, \rm{Hz}$ in each acquisition to avoid $1/f$ noise on the low frequency end and the digital filter roll-off in the high frequency end, while having a large bandwidth to decrease the uncertainty in the mean power density estimation. After computing the median power spectral density we use the aliasing fraction defined in equation \ref{eq:af} for (decimated) sampling rates of  $300.5 \, \rm Hz$ for PA4 and $395 \, \rm Hz$ for PA5 and PA6 (slow) and  $4.8\, \rm kHz$ for PA4 and $5.4\,\rm kHz$ for PA5 and PA6 (fast). 
We estimate the error in the median power spectral density as $\sigma_{mean} = \sigma / \sqrt{N_{bins}}$ where $\sigma$ is the standard deviation taken across frequency bins and $N_{bins}$ is the number of frequency samples in the band. This error in the estimation of the mean power spectral density across the band is $<4\%$ for a single observation, which propagates to a $< 6\%$ uncertainty in the aliasing fraction estimation for each detector individually. Uncertainties can be reduced further assuming detectors of the same frequency (90, 150, $220 \, \rm{GHz}$) have similar noise properties (identically distributed aliasing fractions) and aggregating their statistics.

To discard non-working detectors, we only consider detectors  where the aliasing fraction is higher than 0.5 and lower than 1.5. Aliasing fractions outside this range have a very low probability of belonging to a working detector given the observed distribution as suggested by Monte Carlo simulations \cite{gallardo_optimizing_2019}.

\section{Results}
\label{sec:Results}
\begin{figure}
    \centering
    \begin{minipage}{0.49\textwidth}
        \includegraphics[width=\textwidth]{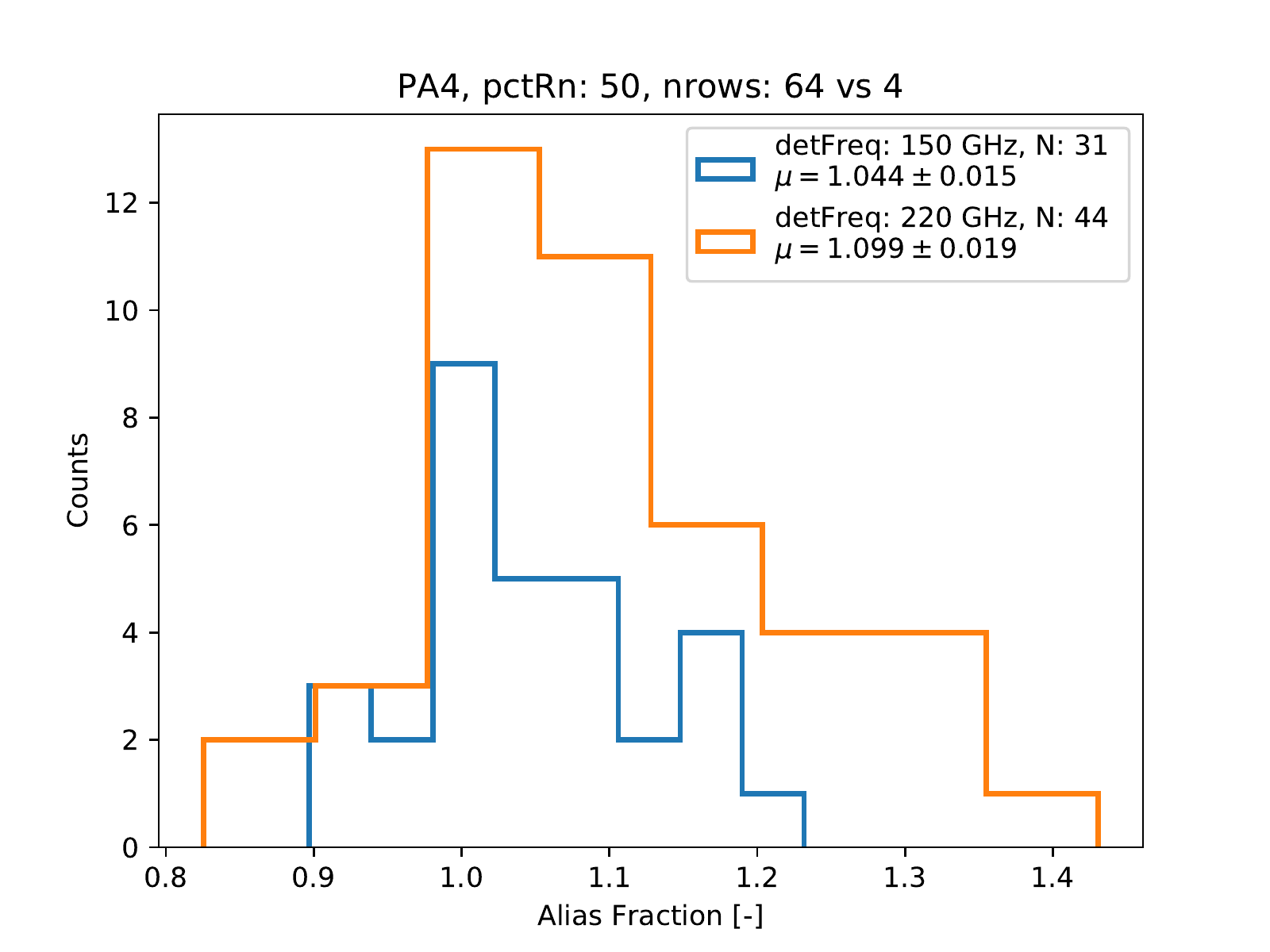}
    \end{minipage}
    \begin{minipage}{0.49\textwidth}
        \includegraphics[width=\textwidth]{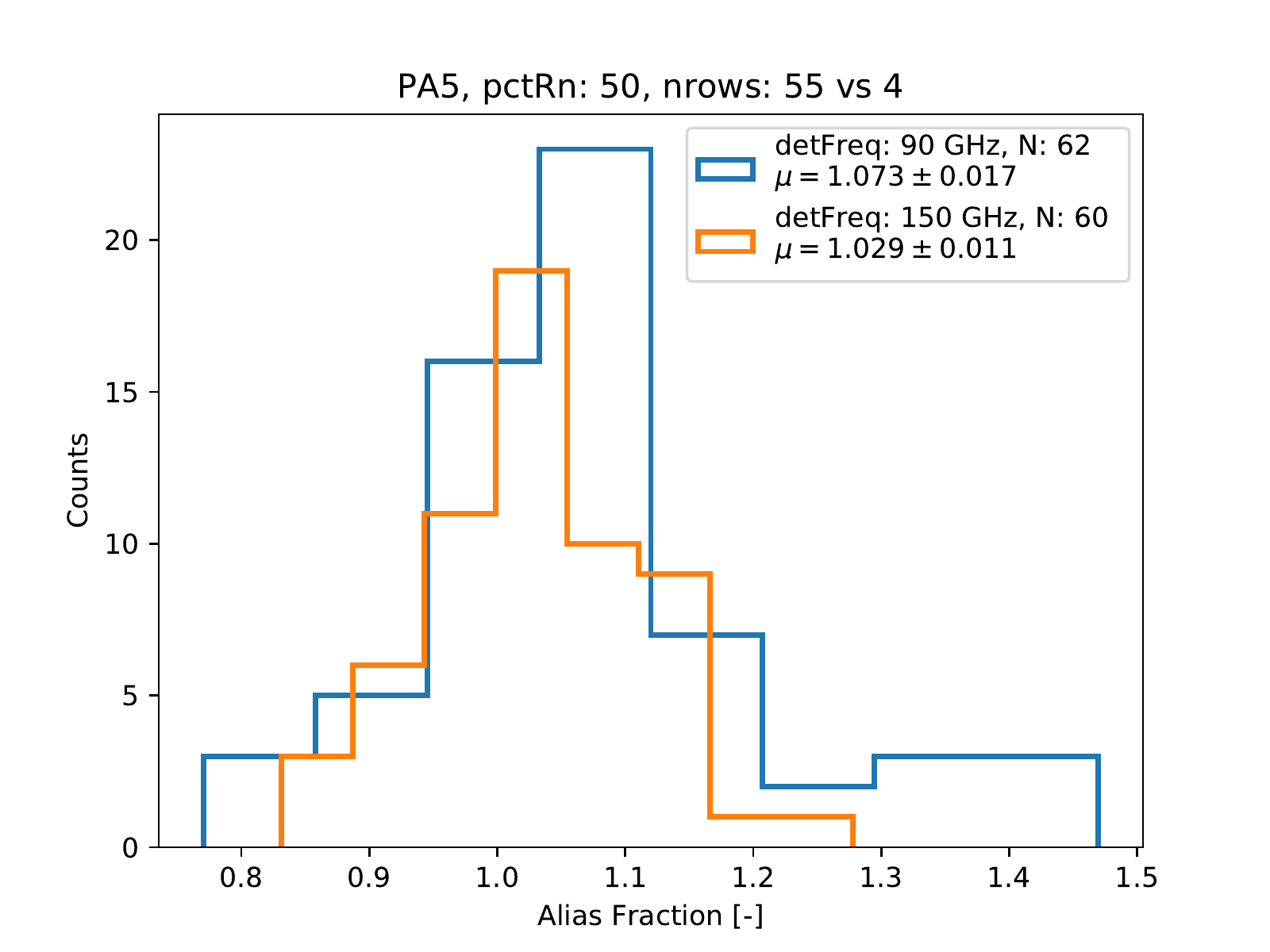}
    \end{minipage}
    
    \begin{minipage}{0.5\textwidth}
        \includegraphics[width=\textwidth]{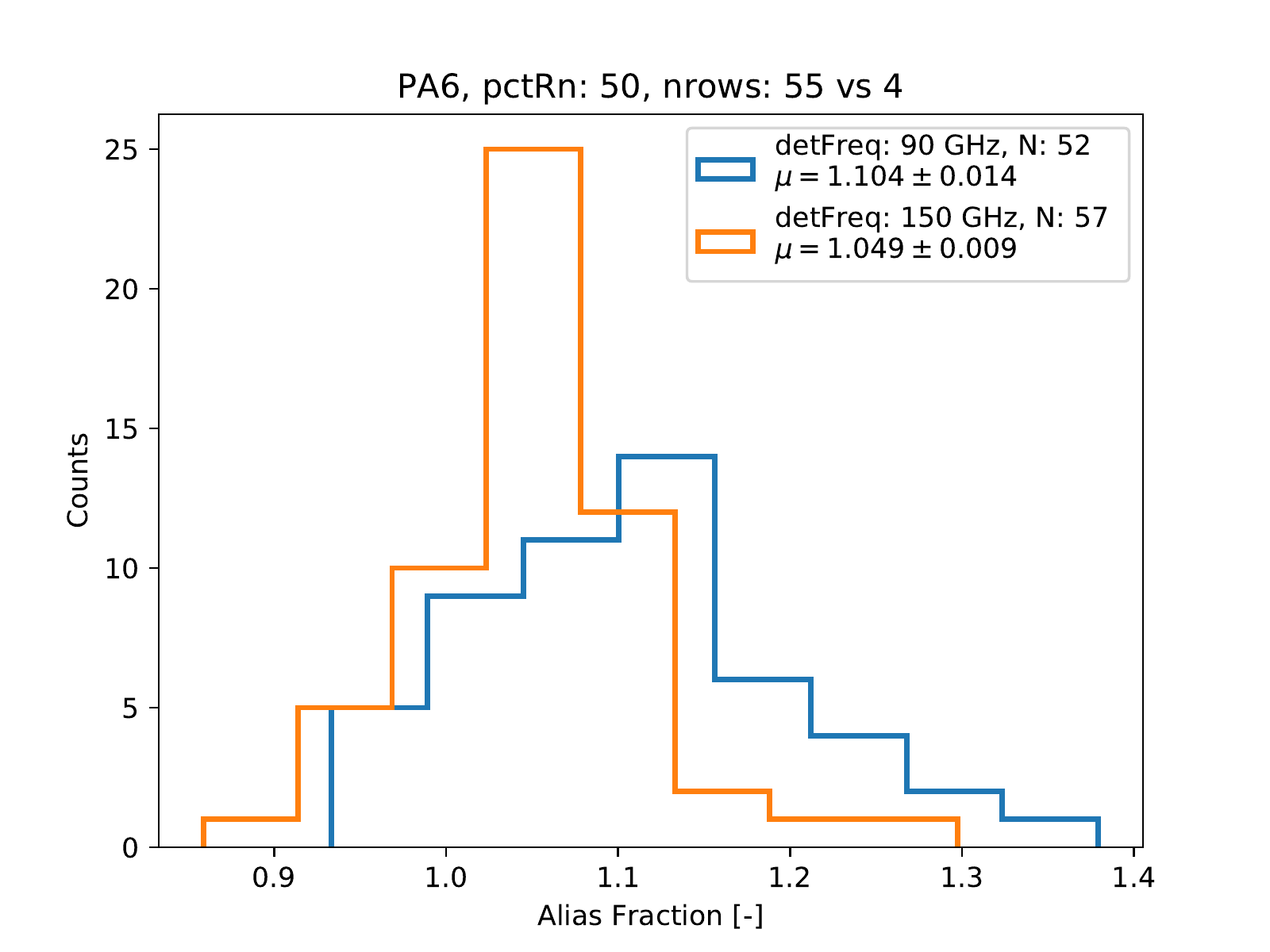}
    \end{minipage}
    \caption{Histograms showing the measured aliasing fraction from three arrays split by detector frequency. Bias point is $50\, \rm{\%R_n}$ and aliasing fraction compares the nominal sampling frequency to the sampling frequency obtained when multiplexing the first 4 rows (see table \ref{tab:Aliasing_mceAquisitionParams}).}
    \label{fig:alising_fraction_histograms}
\end{figure}

\begin{figure}
    \centering
   \includegraphics[width=0.8\textwidth]{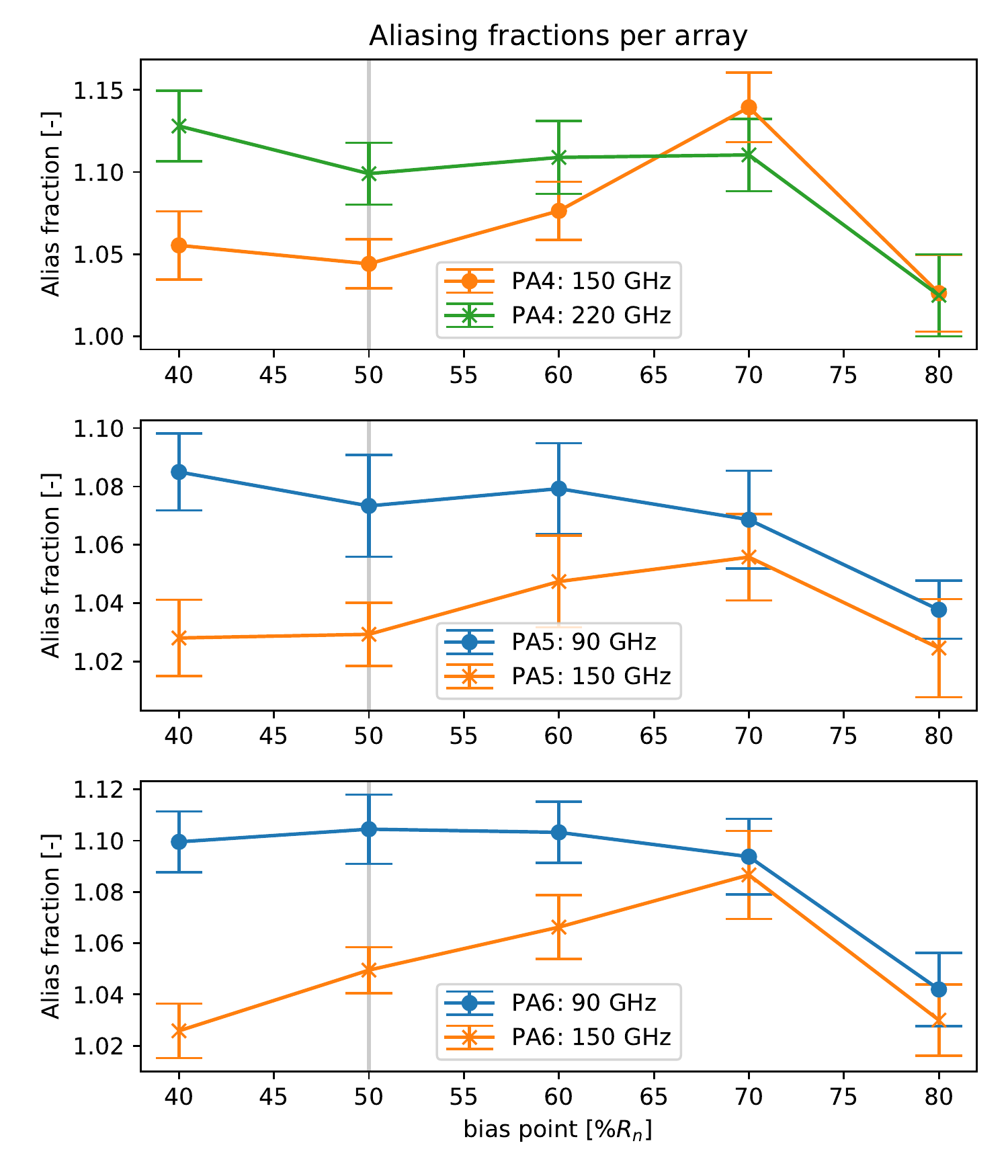}
    \caption{Alias fraction as a function of bias point. Nominal observations target $50\, \rm {\%  R_n}$ (gray vertical line). $150\, \rm{GHz}$ detectors show an increasing aliasing fraction with increasing bias point $R_n$, while $90$ and $220\,\rm{GHz}$ detectors show a flat aliasing fraction from 40 to $70\, \rm{ \%R_n}$.}
    \label{fig:aliasFrac_vs_pctRn}
\end{figure}

After applying the aliasing fraction cut discussed in Section \ref{sec:Measurements} we make histograms of the aliasing fractions for the first four rows of detectors splitting them by detector frequency. Histograms are consistent with the expected distribution from Monte Carlo simulations which use the observed mean power spectral density distribution (discussed in Section \ref{sec:Measurements}) to predict the aliasing fraction distribution. At $50\% \, \rm{R_n}$ we obtain aliasing fractions over unity of 4, 3 and 5$\%$ for the $150\, \rm GHz$ detectors in PA4, PA5 and PA6 respectively. PA4 showed an over unity aliasing fraction of $10 \, \rm \%$ at $220 \, \rm{GHz}$. PA5 and PA6 showed 7 and 10$\, \rm \%$ respectively at 90GHz. Uncertainties are in the 1-2\% level at $1\sigma$. Figure \ref{fig:alising_fraction_histograms} shows aliasing fraction histograms for the three arrays in Advanced ACTPol split by detector frequency at $50\% \,\rm{R_n}$. Labels indicate the detector frequency, the number of detectors that survived the cut and the estimate of the mean of the distribution. Uncertainties are computed as $\sigma/\sqrt{N}$, where $\sigma$ is the aliasing fraction standard deviation across detectors and $N$ is the number of detectors that survived the cut for each detector type.

Figure \ref{fig:aliasFrac_vs_pctRn} shows the mean aliasing fraction estimate as a function of the bias point expressed as a percentage of the normal resistance for the three arrays. Note that the $150\, \rm{GHz}$ detectors show increasing aliasing fractions with increasing point in the transition ($\rm{\%R_n}$). Detectors operating at  $220\, \rm GHz$  and $90 \, \rm GHz$ show a flat response with respect to bias point between 40 and $70 \% \, \rm {R_n}$. Observed trends support the use of a $50\%$ bias point from aliasing fraction estimates, time constants and stability criteria.
\section{Conclusion}
\label{sec:Conclusion}

We present detector noise aliasing measurements to characterize the performance of time-division SQUID multiplexing in the Advanced ACTPol TES arrays. We demonstrate how these measurements can be done in a MCE system by adjusting the readout parameters to read out only a fraction of the array, and thereby increase the sampling frequency. Excess over unity aliasing fractions (in $\left[\frac{dac^2/Hz}{dac^2/Hz}\right]$) are lower than $\sim 10\%$ for the three arrays studied at $50 \, \rm {\%R_n}$. The $150\, \rm {GHz}$ detectors showed aliasing fractions lower than $ 5 \pm 2 \%$.  More precise measurements could be made by increasing the measurement time, which provides more independent realizations of the noise in the system. Photon loading during these measurements was higher than typical observing conditions (we estimate an equivalent precipitable water vapor of $2\, \rm{mm}$ for our covers-on tests), which suggests that the typical aliasing levels during observations are $2$ to $4\%$ higher than presented here.

Time-division SQUID multiplexing is one of the most mature readout approaches for TES arrays and has been adopted as part of the reference design for the next generation CMB-S4 project \cite{2019arXiv190704473A}.  While other projects are implementing different TES readout techniques with unprecedented detector counts that will be considered for CMB-S4, the performance of these alternative approaches will need to be characterized, optimized and compared to benchmarks achieved by the current generation of CMB experiments, like those presented here. 

\begin{acknowledgements}
The authors thank Max Fankhanel and Rodrigo Quiroga for helping with the site operations during the measurements presented here. Comments on this draft were received from two anonymous reviewers.
\end{acknowledgements}

\pagebreak

\bibliographystyle{ltd_style}
\bibliography{references_explicit,ref}

\end{document}